\documentclass[prl,twocolumn,nofootinbib]{revtex4-1}
\usepackage[utf8]{inputenc}
\DeclareUnicodeCharacter{200A}{!!!FIX ME!!!} 
\usepackage{graphicx}
\usepackage{color}
\usepackage{dcolumn}
\usepackage{bm}
\usepackage{amsthm}
\usepackage[bottom]{footmisc}
\usepackage{footnote}
\usepackage{lipsum}
\usepackage{gensymb}
\usepackage{wrapfig}
\usepackage{sidecap}
\renewcommand{\thefootnote}{\fnsymbol{footnote}}

\newcommand\blfootnote[1]{%
\begingroup
\renewcommand\thefootnote{}\footnote{#1}%
\addtocounter{footnote}{-1}%
\endgroup}
\begin{document}





\title{A gate-programmable van der Waals metal-ferroelectric-semiconductor memory}




\author{Wanying Li,$^{1,2*}$ Yimeng Guo,$^{1,2*}$ Zhaoping Luo,$^{1}$ Bo Han,$^{3}$ Xuepeng Zhan,$^{4}$ Kenji Watanabe,$^{5}$ Takashi Taniguchi,$^{6}$ Jiezhi Chen,$^{4}$ Peng Gao,$^{3}$ Xiuyan Li,$^{1}$ Chengxin Zhao,$^{7,8\dagger}$ Zheng Vitto Han,$^{9,10\dagger}$ Hanwen Wang$^{1\dagger}$ }


\affiliation{$^{1}$Shenyang National Laboratory for Materials Science, Institute of Metal Research, Chinese Academy of Sciences, Shenyang 110016, China}
\affiliation{$^{2}$School of Material Science and Engineering, University of Science and Technology of China, Anhui 230026, China}
\affiliation{$^{3}$International Center for Quantum Materials, and Electron Microscopy Laboratory, School of Physics, Peking University, Beijing 100871, China}
\affiliation{$^{4}$School of Information Science and Engineering (ISE), Shandong University, Qingdao, P. R. China}
\affiliation{$^{5}$Research Center for Functional Materials, National Institute for Materials Science, 1-1 Namiki, Tsukuba 305-0044, Japan}
\affiliation{$^{6}$International Center for Materials Nanoarchitectonics, National Institute for Materials Science,  1-1 Namiki, Tsukuba 305-0044, Japan}
\affiliation{$^{7}$Institute of Modern Physics, Chinese Academy of Sciences, Lanzhou 730000, China}
\affiliation{$^{8}$School of Nuclear Science and Technology, University of Chinese Academy of Sciences, Beijing 100049, China}
\affiliation{$^{9}$Collaborative Innovation Center of Extreme Optics, Shanxi University, Taiyuan 030006, China}
\affiliation{$^{10}$State Key Laboratory of Quantum Optics and Quantum Optics Devices, Institute of Opto-Electronics, Shanxi University, Taiyuan 030006, China}


\maketitle
\blfootnote{\textup{*} These authors contribute equally.}

\blfootnote{$^\dagger$Corresponding to: chengxin.zhao@impcas.ac.cn, vitto.han@gmail.com, and hwwang15s@imr.ac.cn}

\textbf{Ferroelecticity, one of the keys to realize non-volatile memories owing to the remanent electric polarization, has been an emerging phenomenon in the two-dimensional (2D) limit. Yet the demonstrations of van der Waals (vdW) memories using 2D ferroelectric materials as an ingredient are very limited. Especially, gate-tunable ferroelectric vdW memristive device, which holds promises in future neuromorphic applications, remains challenging. Here, we show a prototype gate-programmable memory by vertically assembling graphite, CuInP$_{2}$S$_{6}$, and MoS$_{2}$ layers into a metal-ferroelectric-semiconductor architecture. The resulted devices exhibit two-terminal switchable electro-resistance with on-off ratios exceeding 10$^{5}$ and long-term retention, akin to a conventional memristor but strongly coupled to the ferroelectric characteristics of the CuInP$_{2}$S$_{6}$ layer. By controlling the top gate, Fermi level of MoS$_{2}$ can be set inside (outside) of its band gap to quench (enable) the memristive behaviour, yielding a three-terminal gate programmable non-volatile vdW memory. Our findings pave the way for the engineering of ferroelectric-mediated memories in future implementations of nanoelectronics.}


 \bigskip
 \bigskip

   \begin{figure*}[ht!]
   \includegraphics[width=0.95\linewidth]{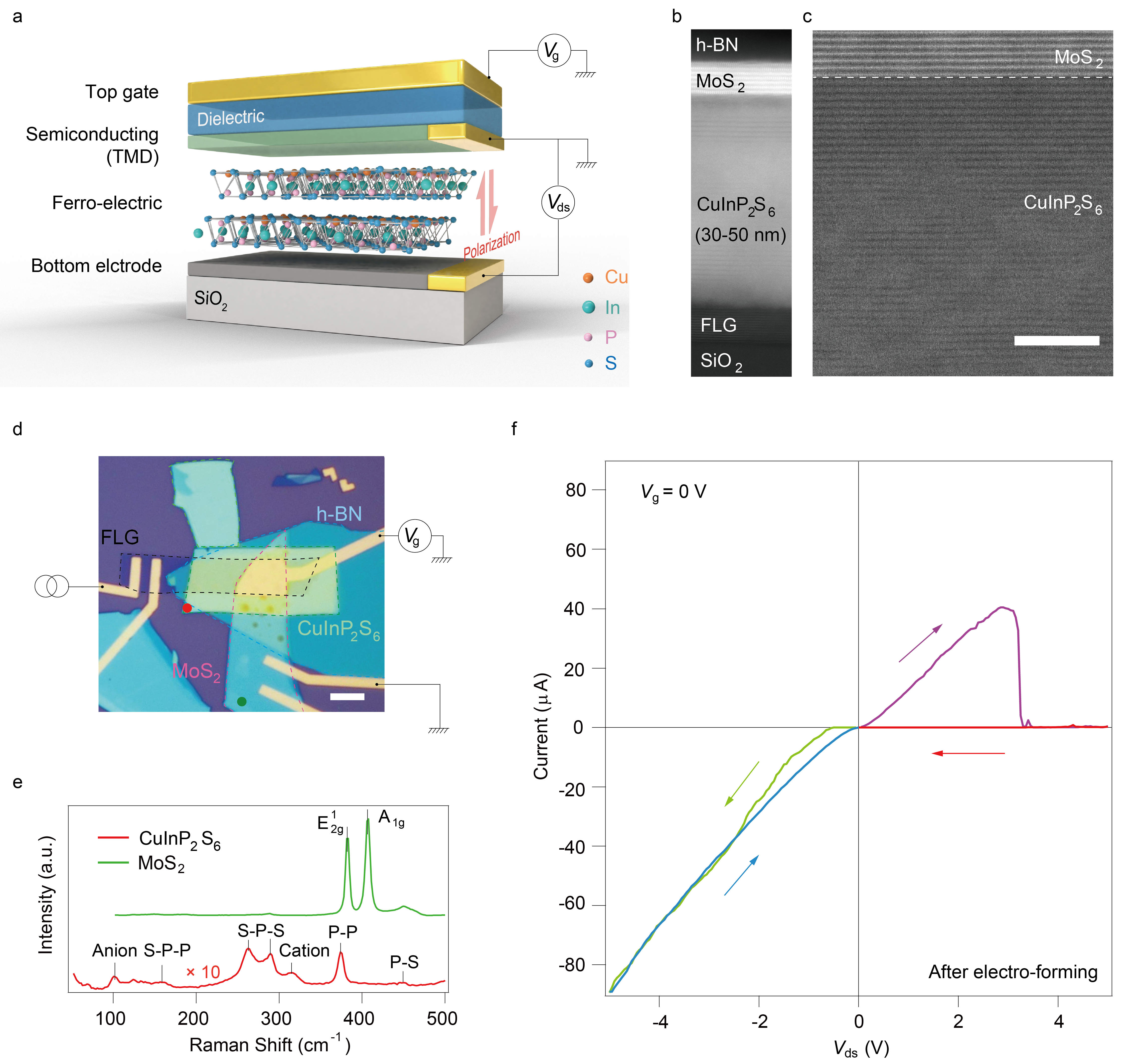}
   \caption{\textbf{Characterizations of vdW M-FE-S heterojunctions.} (a) Art view of the  vdW heterostructure vertically assembled by bottom electrode, ferroelectric, semiconducting, and dielectric layers, respectively, with a Au layer as the top gate. (b) HAADF-STEM image of the cross-section of a typical sample consisted of vdW layers of h-BN, MoS$_{2}$, CuInP$_{2}$S$_{6}$, graphite, in a sequence from top to down.  (c) A zoomed-in view of the atomic structure of the interface of semiconducting MoS$_{2}$ and ferroelectric CuInP$_{2}$S$_{6}$. Scale bar is 10 nm. (d) Optical image of a typical sample in a planar view. Scale bar is 5 $\mu$m. (e) Raman spectroscopy of dotted regions marked in (d). (f) A typical pinched ${I}$-${V}$ curve obtained in a vdW M-FE-S heterojunction, showing characteristic memristive behaviour.
   }
   \end{figure*}

The phenomenon of switchable electro-resistance between on and off states in a metal-insulator-metal (MIM) junction, often facilitated by the formation/deformation of conduction paths through filaments upon application of voltages \cite{Conduction_Filament_Review_2014}, has been an endeavour toward industrial manufacturation of resistive random access memories (ReRAM, deemed also as memristors). Conventionally, such resistive switching behaviours are preferably realized in a two-terminal device in the tunnelling regime, with the insulator in the MIM junction being a high-$\textbf{k}$ dielectric material \cite{2018electronic}. Notably, by substituting the dielectric layer with ferroelectric or multi-ferroic ones, a family of memristors can be established, with their tunneling electro-resistance (TER) inheriting the ferroic-characteristics and exhibiting enhanced tunability of the tunneling barriers \cite{Fert_NM_2007, SolidState_NN_2012, NanoLett_2012, NM_2012, LiQi_NM_2013,  MinNaiben_NM_2013, luo2022high}.

Recently, routes have been developed either to boost the performance or to exploit distinct mechanisms of the memristive devices \cite{zhu2019ionic, huang2020gate}. Among them, vdW devices are of particular promises, as they can be interfaced free of lattice-matching and compatible with a diversity of functionalities \cite{JOS2019, CPB2020}. For example, it is demonstrated that memristive behaviours can be observed in vdW devices in both the tunnelling \cite{CIPS_NE_2020} and the diffusive regimes \cite{Miaofeng_NE_2018, MoS2_NN_2015}. The latter scenarios can be realized with certain mediations (such as ion-migration \cite{Miaofeng_NE_2018}, grain boundaries \cite{MoS2_NN_2015}, and ionic-gate induced phase transition \cite{AM_Synaptic_2018}, while the channels are often laterally arranged with lengths exceeding a few micron meters. The atomically thin lateral semiconducting conduction channel thus enables further gate tunablility \cite{AFM_synaptic_2019}, which leads to multiterminal memtransistors that mimic synaptic responses upon training voltage pulses, holding promises in future neuromorphic computing \cite{AM_Review_2020, sangwan2020neuromorphic, bessonov2015layered, sangwan2018multi}. Besides, compared to lateral memtransistors, vertically assembled memtransistors are more compatible for industrial production thanks to its relatively small footprint \cite{wang2020resistive}. To date, studies of vdW memtransistors however remain limited, and, especially, vertically assembled gate-tunable ferroelectric vdW memristive device has been missing. 

In this work, we utilize the vdW vertical assembly as a platform to devise a prototype gate-programmable memory by stacking few-layers of graphene, CuInP$_{2}$S$_{6}$, and MoS$_{2}$  into a metal-ferroelectric-semiconductor (M-FE-S) architecture, with the top semiconducting layer simultaneously equipped into a metal-oxide-semiconductor field effect transistor (MOS-FET). The M-FE-S devices show resistive switch with stability and on-off ratios reaching 10$^{5}$. Interestingly, the obtained memristive characteristics can be enabled and disabled using the top gate at room temperature, and can be killed by heating the system above the ferroelectric transition Curie temperature of CuInP$_{2}$S$_{6}$. We attribute this phenomenon to a ferroelectric field effect at the FE-S interface. The proposed three-terminal gate programmable non-volatile vdW M-FE-S memory can, in principle, be trained into artificial synapses, shedding light on the design of ferroelectric-mediated memories in future nanoelectronics.

\section{Results}
\textbf{Fabrication and characterizations of M-FE-S memristors.} A vdW vertical architecture illustrated in Fig. 1a is adopted, to construct an M-FE-S junction in the lower part and a MOS-FET in the upper part. In this configuration, ferroelectic-semiconducting interface as well as a gate tunable semiconducting channel can be combined in one single device, which is essential for the realization of a gate-programmable M-FE-S memory, as will be discussed in the following sections. By adopting the dry-transfer method \cite{Lei_Science}, multi-layered vdW heterostructure described in Fig. 1a was fabricated by stacking few-layers of graphene, CuInP$_{2}$S$_{6}$, MoS$_{2}$, and hexagonal boron nitride (h-BN), with a graphite layer (5-10 nm in thickness) serving as the bottom electrode (see Supplementary Figure S1). Atomic resolution of the cross section of a typical heterostructure can be seen in the high-angle annular dark-field scanning transmission electron microscopy (HADDF-STEM) image in Fig. 1b, with the corresponding zoomed-in view of the FE-S interface shown in Fig. 1c. It is seen that the  layered structure of CuInP$_{2}$S$_{6}$ and MoS$_{2}$ are well defined, and the FE-S interface is atomically sharp. Optical micrograph of a typical sample is shown in Fig. 1d, with each constituent layer highlighted by different coloured dashed lines. Electrodes and top gates were patterned via standard lithography and electron-beam evaporation. Raman spectroscopy of CuInP$_{2}$S$_{6}$ and MoS$_{2}$ layers were measured to confirm the phonon modes of each crystal, as shown in Fig. 1e. It is seen that both of the two materials exhibit consistent Raman peaks as compared to the previous reported results \cite{si2018ferroelectric, li2012bulk}. It is noteworthy that the anion and cation peaks can be found in the Raman spectrum of CuInP$_{2}$S$_{6}$, indicating the existence of ferroelectric dipole polarization at room temperature in the flake. Indeed, Curie temperature $T_\mathrm{C}$ of few layered CuInP$_{2}$S$_{6}$ is around 315 K \cite{LiuZheng_NC_2016}. 

In order to effectively tune the carrier density of the semiconducting layer (few-layered MoS$_{2}$, for example), the underneath ferroelectric layer has to be thick enough to avoid leakage and to allow a well defined chemical potential in the MoS$_{2}$ layer. It is known that ferroelectric tunnelling memory was reported recently \cite{CIPS_NE_2020}, where a 4-nm-thick CuInP$_{2}$S$_{6}$ layer was used to form a simple M-FE-M tunnel junction. However, as mentioned above, such structure cannot be compatible with gate-tunability. Therefore, CuInP$_{2}$S$_{6}$ flakes with thicknesses of 30 $\sim$ 50 nm were used to fabricate the M-FE-S junctions in the current study to retain their dielectric properties.

   \begin{figure}[ht!]
   \includegraphics[width=0.95\linewidth]{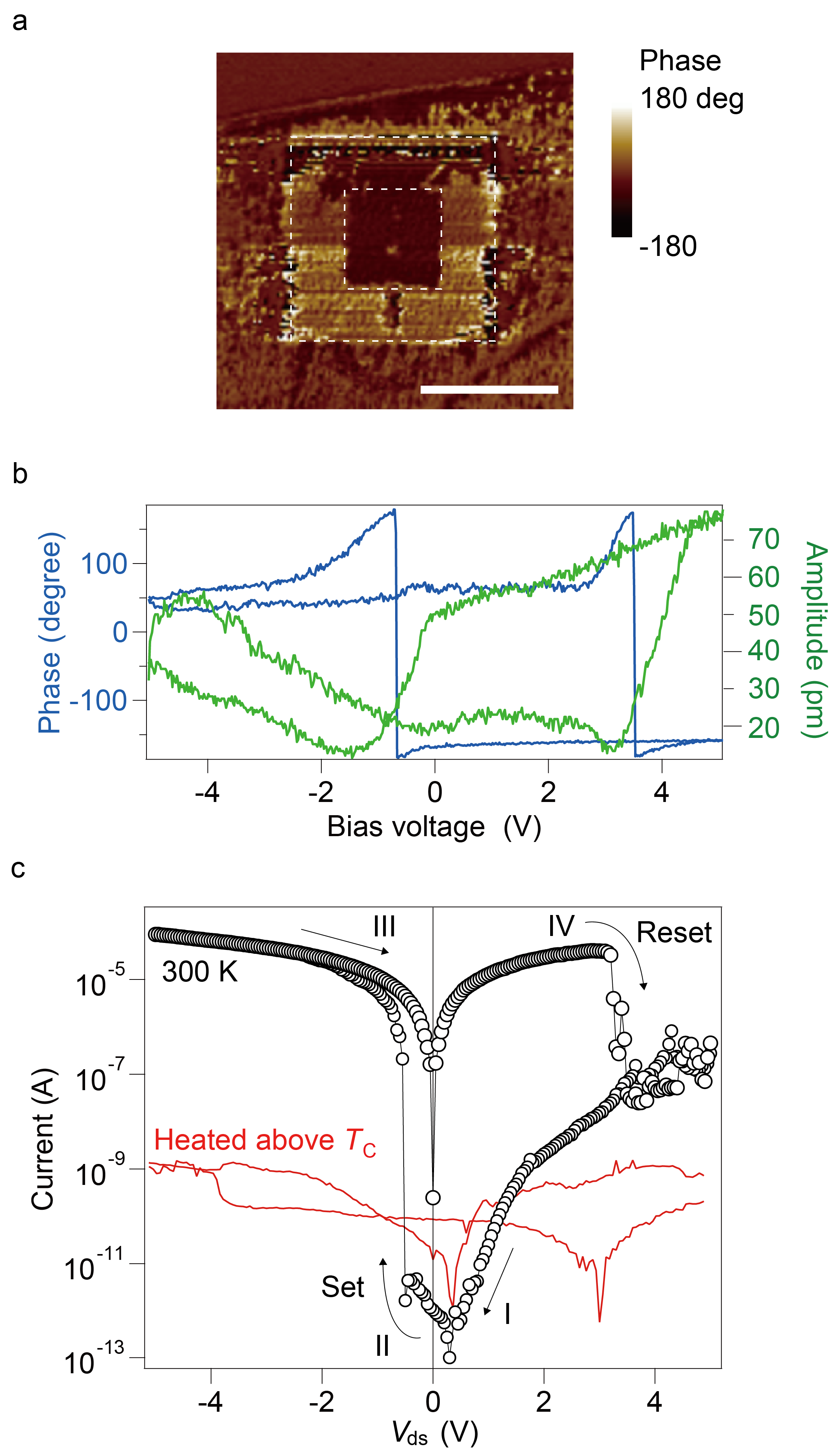}
	   \caption{\textbf{Ferroeletricity-assisted resistive switching behaviour in vdW M-FE-S heterojunctions.} (a) PFM phase images for a typical CuInP$_{2}$S$_{6}$ flake (thickness $t \sim$ 40 nm) with written box-in-box pattern with reverse DC bias.Scale bar is 5 $\mu$m. (b) Out-of-plane amplitude (green) and phase (blue) measurements obtained by PFM on a typical CuInP$_{2}$S$_{6}$ flake with $t$ $\sim$ 40 nm. (c) Looped $I$-$V$ characteristics in a log-scale showing typical resistive switching behaviour, as presented by the curve of open circles. Directions of the sweep are indicated by arrows, with four distinct stages. The device is reset to HRS during stage IV at certain positive $V_\mathrm{ds}$, and set to LRS during stage II at certain negative $V_\mathrm{ds}$. The device show absence of memristive performance when heated above $T_\mathrm{C}$, as shown in the red curve.}
   \label{fig:fig2}
   \end{figure}

We now characterize the two-terminal electro-resistance in the as-prepared M-FE-S junctions. During the electrical measurements, the MoS$_{2}$ layer was kept grounded, while bias voltages were applied on the graphite (or few-layered graphene, FLG) layer. As shown in Fig. 1f, looped $I$-$V$ characteristics of a representative device is plotted, with the sweep-directions indicated by arrows. This pinched $I$-$V$ loop is a typical resistive switch curve, with four distinct stages. Specifically, the device is switching between a high resistance state (HRS) and a low resistance state (LRS) with a hysteresis that is depending on the sweeping direction and maximum polarization voltage. Multiple devices exhibit consistent behaviour, as shown in Supplementary Figure S2. We attribute this resistive behaviour to the formation of conduction paths in the CuInP$_{2}$S$_{6}$ layer, as the devices all undergo an electro-forming process before they can serve as memristors, shown in Supplementary Figures S3.

\begin{figure}[ht!]
\includegraphics[width=0.9\linewidth]{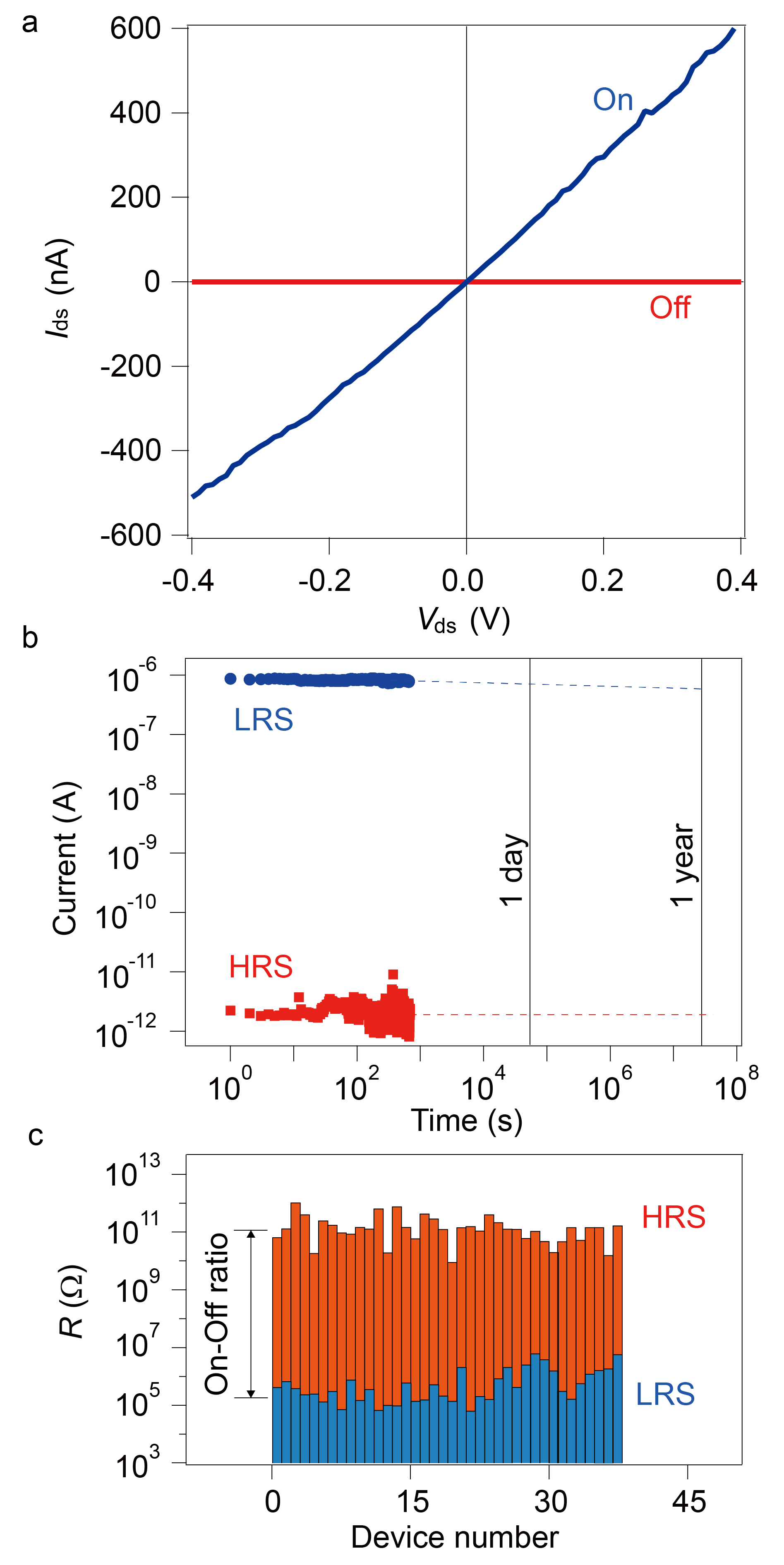}
\caption{\textbf{Performances of two-terminal electro-resistances in vdW M-FE-S heterojunction memristors.} (a) $I$-$V$ curve for the same device in the ON and OFF states. (b) Retention time of a typical device measured at $V_\mathrm{read}$ = 0.5 V after applying a 100 $\mu$s voltage pulse of $V_\mathrm{ds}$=4.0 V (-4.0 V) to set the OFF (ON) state. (c) Statistics of HRS and LRS states measured in more than 30 different samples. Voltage pulses of 100 $\mu$s were used to set the OFF (ON) states. Average ON/OFF ratio is estimated to be about 10$^{5}$.}

\label{fig:fig3}
\end{figure}

\textbf{Ferroelectric characteristics of the M-FE-S memristor.} In the following, we demonstrate that the fabricated M-FE-S memristors are strongly affected by the ferroelectric properties of the few-layered CuInP$_{2}$S$_{6}$. Firstly, it's worth mentioning that the electrical resistance of the M-FE-S junctions might be dominated by the Schottky barrier at the FE-S interface, which is carefully discussed in the Supplementary information Figure S4. Piezoresponse force microscopy (PFM) is used to characterize the ferroelectricity of the CuInP$_{2}$S$_{6}$ flakes. As shown in Fig. 2a, using a PFM tip with reverse DC bias voltages, ferroelectric domain of a box-in-box pattern was written and scanned in the PFM phase image, and the pattern can be reserved for more than one year, as shown in Supplementary Figure S5. Furthermore, out-of-plane amplitude (green) and phase (blue) obtained on a typical CuInP$_{2}$S$_{6}$ flake with $t$ $\sim$ 40 nm are given in Fig. 2b, exhibiting hysteresis loops with asymmetric coercivity fields biased toward positive voltage. Such biased polarization-switching fields are also seen in previous reports \cite{LiuZheng_NC_2016, chen2019thickness,brehm2020tunable}. We notice that for Cr/CuInP$_{2}$S$_{6}$/FLG devices, there is no distinct resistive switching behaviour in the diffusive regime, as shown in Supplementary Figure S6. Besides, negligble resistive switching behaviour is present in other similar M-FE-M structures, e.g. Au/CuInP$_{2}$S$_{6}$/Au and Cr/CuInP$_{2}$S$_{6}$/Au, indicating that the resistive switching behaviour is related to the FE-S interface, shown in Supplementary Figure S7-8.

Interestingly, the switching voltages (about -1.0 V and +2.5 V for each polarization directions) in CuInP$_{2}$S$_{6}$ in Fig. 2b are inherited by the resistive switching $I$-$V$ curve in the M-FE-S memristor with similar thickness of CuInP$_{2}$S$_{6}$, as plotted in a log-scaled looped $I$-$V$ in Fig. 2c (as typical linear $I$-$V$ data is shown in Fig. 1f). Four distinct stages along with the sweeping-up and down directions are indicated by arrows in Fig. 2c. At the vicinity of positive switching voltage ($V_\mathrm{ds}$) of the peizoelectric loop in Fig. 2b, the M-FE-S memristor device is ``reset'' to HRS during stage IV. On the contrary, the device is ``set'' to LRS during stage II at the vicinity of negative switching voltage. It is worthy note that the voltage differences between ``set'' and  ``reset'' coincidence well with the differences between polarization-switching voltages of CuInP$_{2}$S$_{6}$ \cite{chen2019thickness}, as shown in the Supplementary information Figure S9. Mechanism of the observed ferroelectricity-assisted FE-S memristive behaviour can be attributed to both the formation of conduction paths in the CuInP$_{2}$S$_{6}$ layer, and a modulation of built-in ferroelectrical field at the FE-S interface and lead to an effectively depleted states in the semiconducting layer, leading to a tunable interfacial barrier induced by the ferroelectric polarization \cite{MinNaiben_NM_2013}.

Moreover, since the $T_\mathrm{C}$ of the few-layered CuInP$_{2}$S$_{6}$ studied in this work is about 315 K, it is crucial to test the M-FE-S memristors above $T_\mathrm{C}$. As expected, the resistive switching of the same device is quenched when heated at 400 K, as shown in the red looped $I$-$V$ curve in Fig. 2c. Even when cooled to room temperature, no memristive behaviour was observed, suggesting that the memristive behaviour of the M-FE-S device is strongly coupled to the ferroelectric nature of CuInP$_{2}$S$_{6}$, since ferroelectric domains become randomly distributed when the system is naturally cooled down from above $T_\mathrm{C}$. The resistive switching thus may require a ferroelectric domain training process in the system, similar to the initial magnetization process in some hard ferromagnets \cite{broadway2020imaging}. Indeed, by re-doing the electro-forming process, the absence of memristive performance can be recovered in those annealed samples, as illustrated in Supplementary Figure S10.

\begin{figure*}[ht!]
\includegraphics[width=0.95\linewidth]{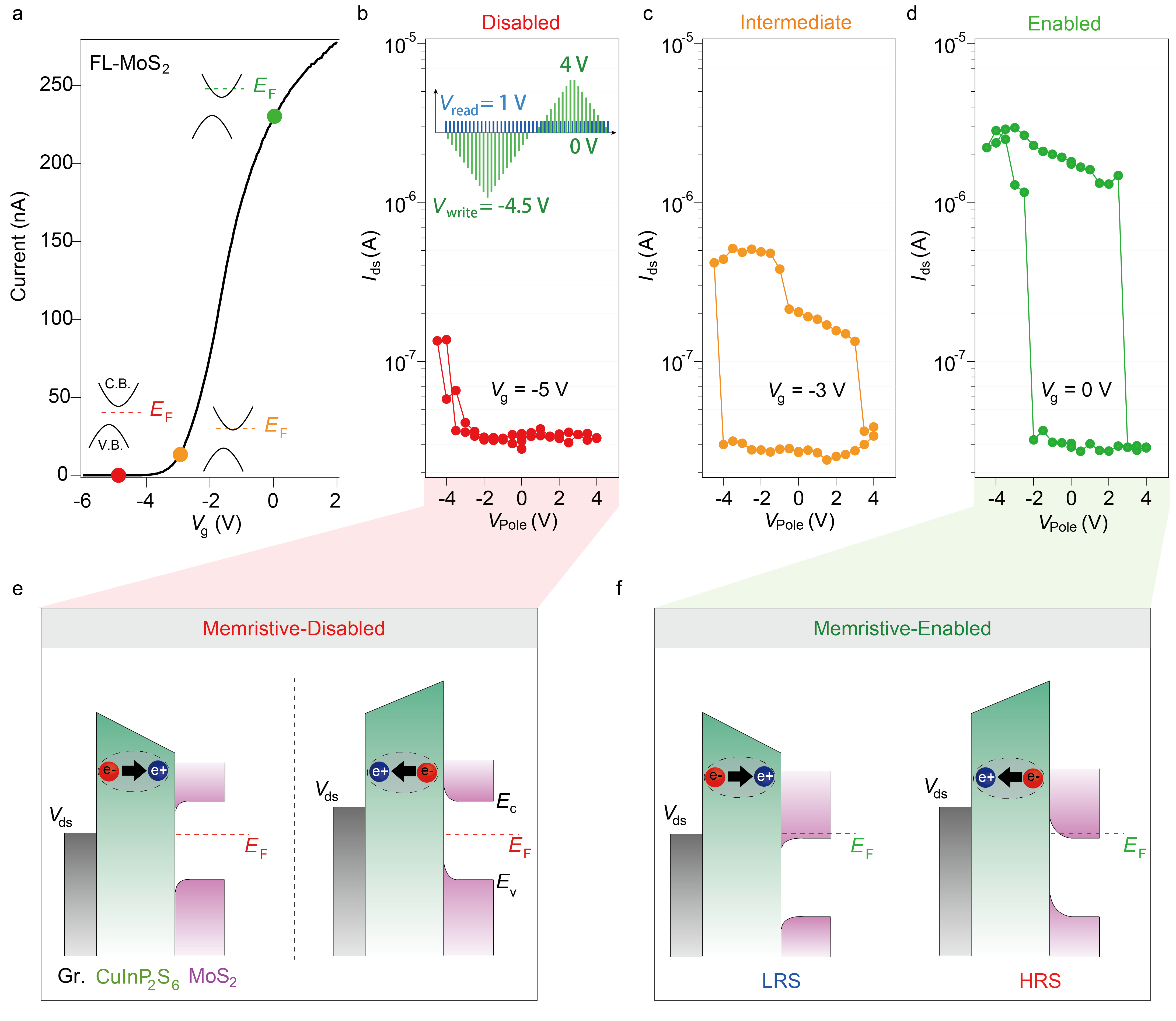}
\caption{\textbf{Multi-level ``ON'' states of vdW M-FE-S gate-programmable memory at room temperature.} (a) Field effect curve of few-layered MoS$_{2}$, measured on a typical MoS$_{2}$ flake with similar thickness in the tested M-FE-S memristor. Illustration of conduction and valence bands with the Fermi energy were given, for three typical states. $E_\mathrm{F}$ denotes the Fermi level. (b)-(d) The memory windows (i.e., Resistance versus pulse-voltage loop) measured at different Fermi energies of the MoS$_{2}$ layer, set by the $V_\mathrm{g}$. Traces of the training pulses of polarization voltage $V_\mathrm{Pole}$ is illustrated in the inset of (b). (e)-(f) Band alignments of the metal-ferroelectric-semiconductor heterojunction. The memristive-disabled and memristive-enabled regimes are realized by adjusting the Fermi energy MoS$_{2}$ inside the gap, and in the electron side, respectively. In the scenario of (f), the HRS and LRS are then achieved via the ferroelectric field effect at the FE-S interface of CuInP$_{2}$S$_{6}$/MoS$_{2}$.}

\label{fig:fig4}
\end{figure*}

\textbf{Performances of two-terminal electro-resistances in vdW M-FE-S memristors.} Figure 3a shows two-terminal $I$-$V$ curves between -0.4 and 0.4 V in the ON and OFF states. Both are nonlinear, and consecutively tuned from negative to positive $V_\mathrm{ds}$ for each states, which differ from the grain-boundary mediated resistive switching devices \cite{MoS2_NN_2015}, but resembles the behaviour of other TER memristors \cite{kim2012ferroelectric}. It is noticed that the devices may be permanently damaged due to dielectric breakdown at $V_\mathrm{ds}$ higher than 6.0 V (more details can be found in Supplementary Figure S11).

Figure 3b shows resistance versus time of a two-terminal M-FE-S device measured at $V_\mathrm{read}$ = 0.5 V after applying a 100 $\mu$s voltage pulse of $V_\mathrm{ds}$=4.0 V (-4.0 V) to set the OFF (ON) states. The devices show retention time which can be extrapolated to the time scale of years, with the attenuation of ON/OFF ratio less than one order of magnitude. By measuring HRS and LRS states in more than 30 different samples (data collected with setting-resetting voltage pulses of 100 $\mu$s with amplitudes of $\pm$4.0 V) are shown in the statistics in Fig. 3c. It is seen that good reproducibility can be obtained in all tested samples, with an average ON/OFF ratio of about 10$^{5}$. It is noticed that among the measured M-FE-S samples, thicknesses of the ferroelectric and semiconducting layers are in the range of 30 to 50 nm, and 5-15 nm, respectively. The variation of thicknesses among samples may be a cause of the fluctuation of ON/OFF ratios observed in Fig. 3c.

\section{Discussion}

To this stage, we have demonstrated the subpart of the as-prepared vertical vdW heterojunction, i.e., the M-FE-S junction that serves as a two-terminal memristor. Now we come to the discussion of the MOS-FET equipped on the top part of the heterojunction illustrated in Fig. 1a. One of the key role of the architecture adopted here is that the vdW semiconductor in the M-FE-S junction can be gated to tune the carrier density (or, the Fermi level) in itself, further affecting the memristive performance as a consequence. As shown in Fig. 4a, field effect in the MoS$_{2}$ few-layers in a typical sample is confirmed. The threshold voltage of the ON-state of the FET is about -4 V, below which the MoS$_{2}$ is tuned into gapped region where no density of states is available. The general FET behaviour of few-layered MoS$_{2}$ crystals in this work is in agreement with previously reported results \cite{li2017gate}. It is noticed that the threshold voltages of the ON-state of MoS$_{2}$ flakes with thickness around 10 nm are similar, as shown in Supplementary information S12.

Three representative gate voltages were chosen to pin the Fermi level of MoS$_{2}$ at the position of inside the conduction band (C.B.), at the bottom of C.B., and in the band gap (between C. B. and the valence band, V. B.), as indicated by green, orange, and red cartoons along with corresponding dots in the field effect curve in Fig. 4a. Then the memory windows (i.e., resistance versus pulse-voltage loop) of the M-FE-S were measured, with $V_\mathrm{g}$ fixed at the above three different values. During the measurement of memory windows, training pulses of polarization voltage $V_\mathrm{Pole}$ were fed in a sequence of each pulse of $V_\mathrm{Pole}$ arranged in a triangle form with each pulse spaced by a constant reading voltage $V_\mathrm{read}$ pulse of 1.0 V, illustrated in the inset of Fig. 4b. It is seen that, by tuning the $V_\mathrm{g}$, the memristive characteristics of the device can be programmed into disabled-, intermediate-, and enabled-states, as shown in Fig. 4b-d, respectively. 

Taking the quenched (Fig. 4b) and restored (Fig. 4d) memristive switching cases as examples, band alignments of the metal-ferroelectric-semiconductor heterojunction are illustrated in Fig. 4e and 4f, respectively, to address the mechanism of the observed gate-programmable features in the memristors. When the Fermi level of the semiconducting MoS$_{2}$ layer is set into the gapped region, the density of states are diminished between the chemical potential window of source and drain, thus the device exhibits negligible conductance in general and the memristive behaviour is quenched, regardless of the direction of ferroelectric polarizations in the M-FE-S structure, shown in Fig. 4e. When the Fermi level of the semiconducting MoS$_{2}$ layer is set into the conduction band, the electrons can be transported between source and drain, leading to the restore or enable of the memristive performance. In this scenario, the direction of ferroelectric polarizations in the M-FE-S structure will effectively form a band bending and lead to a extra depletion in the MoS$_{2}$ when the ferroelectric polarization is pointing against the semiconducting layer, setting the system to the HRS, and vice-versa \cite{MinNaiben_NM_2013}. The gate-programmable nature of the as-prepared M-FE-S memories is key for future engineering of such as artificial synapses and may trigger the exploit of exotic functionalities based on ferroelectric 2D materials. Furthermore, the picture of principle-of-work in our devices can be further expanded into vertical vdW heterojunctions with a broader family of 2D materials.

\bigskip

To conclude, we devised a conceptual vdW vertical architecture with an M-FE-S junction in the lower part and a MOS-FET in the upper part, experimentally constructed by sequentially stacking few-layers of graphene, CuInP$_{2}$S$_{6}$, MoS$_{2}$, and hexagonal boron nitride (h-BN) from bottom to top. In this manner, the conjugated nanostructure can pack memristor and MOSFET into one single device, as the fine-tuning of Fermi level in the semiconducting layer can simultaneously affect the resistive switching behaviour of the M-FE-S junction. When performed as conventional two-terminal memristors, the M-FE-S devices show resistive switch with on-off ratios reaching 10$^{5}$ and state-of-the-art retention time for information storage. The obtained memristive characteristics are strongly coupled to the ferroelectric nature of CuInP$_{2}$S$_{6}$, and can be further programmed, using the top gate at room temperature, into quenched- or enabled-states. The resulted three-terminal gate programmable non-volatile vdW M-FE-S memory is a potent platform for the design of future ferroelectric-mediated vdW memories, which may find applications in such as neuromorphic computing.

\section{Methods}

\textbf{Sample fabrication.} vdW few-layers of the M-FE-S junctions were obtained by mechanically exfoliating high quality bulk crystals. The vertical assembly of vdW layered compounds were fabricated using the dry-transfer method in a nitrogen-filled glove box. Electron beam lithography was done using a Zeiss Sigma 300 SEM with an Raith Elphy Quantum graphic writer. Top gates as well as contacting electrodes were fabricated with an e-beam evaporator, with typical thicknesses of Ti/Au $\sim$ 5/50 nm.

\textbf{TEM characterizations.} TEM characterizations were carried out on a double aberration corrected FEI Themis G2 60-300 electron microscope equipped with a SuperX-EDS detector and operated at 300 kV. Cross sections of as-prepared devices were made using a focused ion beam tool of FEI Helios PFIB CXe cut at 30 kV and polished at 30 and 12 kV.

\textbf{Morphology tests.} A Bruker Dimension Icon AFM was used for thicknesses and morphology tests, as well as PFM characterizations. Optical images were collected by a Nikon LV-ND100 microscope. 

\textbf{Electrical measurements.} The high precision of current measurements of the devices were measured using a Cascade M150 probe station at room temperature, with an Angilent B1500A Semiconductor Device Parameter Analyzer. Gate voltages on the as-prepared multi-terminal memristors were fed by a Keithley 2400 source meter. During measurements, the TMD layer was kept grounded, while the $V_\mathrm{ds}$ was applied on the graphite bottom electrode. It is noticed that if the configuration of source-drain is reversed, the $I$-$V$ characteristics will be reversed accordingly (Supplementary Figure S13). For the memory window measurement in Fig. 4 in the main text, a pulse train constains varying program pulse and constant read pulse (1.0 V) is adopted by using the waveform generator fast measurment unit (WGFMU).The program voltage ranges from from -4.5 V to 4.0 V at a step of 0.5 V.The pulse duration is 50 ms and interval is 10 ms.

\section{\label{sec:level1}DATA AVAILABILITY}
The data that support the findings of this study  will be available at the open-access repository Zenodo (2022) with a doi link, when accepted for publishing.

\section{\label{sec:level2}Code AVAILABILITY}
The computational codes that support the findings of this study are available from the corresponding authors upon reasonable request.

\section{\label{sec:level3}ACKNOWLEDGEMENT}
This work is supported by the National Key R$\&$D Program of China (No. 2019YFA0307800) and the National Natural Science Foundation of China (NSFC) (Grant Nos. 12004389, 12104462, 11974357, U1932151, and 11804386).

\section{Author contributions}
Z.H., H.W., and C.Z. conceived the experiment and supervised the overall project. W.L., Y.G., and H.W. fabricated the devices and carried out electrical transport measurements; K.W. and T.T. provided high quality h-BN bulk crystals. H.W. and Z.H. analysed the data, with W.L., Y.G., and C.Z. participated in. Z.L., X.L., B.H., and P.G. performed TEM characterizations. X.Z. and J.C. carried out resistance versus pulse-voltage loop measurements. The manuscript was written by Z.H., H.W., and W.L. with discussion and inputs from all authors.

\section{ADDITIONAL INFORMATION}
Competing interests: The authors declare no competing financial interests.

\bibliography{Ref_V1}
\bibliographystyle{naturemag}

\end{document}